\documentclass[aps,pra,twocolumn,floatfix,floats,showpacs,superscriptaddress]{revtex4}
\usepackage{graphicx,latexsym}
\usepackage{dcolumn}
\usepackage{amsmath}

\def\eq#1{Eq.~(\ref{#1})}

\begin{document}
\title {Differential conductance of a saddle-point constriction\\
with a time-modulated gate-voltage}
\author{C. S. Tang}
\affiliation{Physics Division, National Center for Theoretical
        Sciences, P.O.\ Box 2-131, Hsinchu 30013, Taiwan}
\author{C. S. Chu}
\affiliation{Department of Electrophysics, National Chiao Tung
University, Hsinchu 30010, Taiwan}
\begin{abstract}
The effect of a time-modulated gate-voltage on the differential
conductance $G$ of a saddle-point constriction is studied.  The
constriction is modeled by a symmetric saddle-point potential and
the time-modulated gate-voltage is represented by a potential of the
form $V_{0}\,\theta(a/2-|x-x_{c}|)\,\cos (\omega t)$. For
$\hbar\omega$ less than half of the transverse subband energy level
spacing, gate-voltage-assisted (suppressed) feature occurs when the
chemical potential $\mu$ is less (greater) than but close to the
threshold energy of a subband.  As $\mu$ increases, $G$ is found to
exhibit, alternatively, the assisted and the suppressed feature. For
larger $\hbar\omega$, these two features may overlap with one
another. Dip structures are found in the suppressed regime.
Mini-steps are found in the assisted regime only when the
gate-voltage covers region far enough away from the center of the
constriction.
\end{abstract}
\maketitle

\section{Introduction}

The effects of time-modulated fields on the quantum transport have
been of continued interest in the recent past. These time-modulated
fields can be transversely
polarized\cite{h1,h3,F3,w3,f3,j4,g4,g5,c6,m6}, longitudinally
polarized\cite{c4,w6}, or represented by time-modulated potentials,
with no polarization\cite{B2,C5,j0,a1,b2,r3,w5,t6,ta6}.  The systems
recently considered are mostly mesoscopic systems, such as the
narrow constrictions\cite{h1,h3,F3,w3,f3,j4,g4,g5,c6,m6,b2,ta6}. For
the case when a constriction is acted upon by an incident
electromagnetic wave, the time-modulated field has a polarization.
This situation can be realized experimentally, as is demonstrated by
two latest experiments\cite{w3,j4}.  On the other hand, the
time-modulated potentials are expected to be realized in
gate-voltage configurations\cite{go5,ta6},  which is shown in Fig.\
1.
\begin{figure}[btp]
      \includegraphics[width=0.40\textwidth,angle=0]{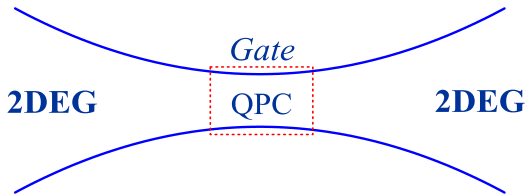}
      \caption{Sketch of the gated saddle-point constriction which
            is connected at each end to a two-dimensional
            electron gas electrode.  The gate induces a
            finite-range time-modulated potential in the
            constriction.}
\end{figure}

The presence of the time-modulated fields, with or without
polarizations, gives rise to coherent inelastic scatterings.
These inelastic scatterings do not conserve
the longitudinal momentum along the transport direction,
as long as the time-modulated fields have finite longitudinal ranges.
The reason being that the finiteness in the range of the
fields breaks the translational invariance
\cite{c6,ta6}.  Furthermore, the inelastic scattering processes
induced by these time-modulated fields depend also on the polarization
of the fields.  In an adiabatically varying constriction, the
inelastic scattering processes involve inter-subband transitions, when
the time-modulated fields are transversely polarized, but involve only
intra-subband transitions, when the fields do not have polarizations,
such as those arise from time-modulated gate-voltages.  The
detail transport characteristics of the constriction hence depend
on the polarization of the time-modulated fields.

In this work we focus on the case of a time-modulated potential. The
effect of such a potential on the transport properties of one
dimensional systems have been investigated in many previous
works\cite{B2,C5,j0,a1,b2,r3,w5,t6,ta6}.  However, the possible
manifestation of quasi-bound-state (QBS) features has not been
widely recognized, except for the work of Bagwell and Lake
\cite{b2}, who have considered a time-dependent potential that has a
delta profile.  The energy of this QBS is below, but close to, the
band bottom of the one dimensional system.  The transport exhibits
QBS feature when the conducting electrons can make transitions via
inelastic processes to the QBS. We expect this QBS feature to be
more significant in a narrow constriction than in a one dimensional
system, because there are, in a constriction, more  subbands and
hence more QBS's. Furthermore, the tunability of the subband
structures and the chemical potential together provide greater
feasability for probing the QBS feature in narrow constrictions.

In an earlier work\cite{ta6}, the present authors have investigated
the effect of a time-dependent gate potential acting upon the
uniform-width region of a narrow constriction. For those electrons
that manage to enter the narrow channel region from the two
end-electrodes, they have perfect transmission, and the
time-modulated potential cannot further increase the dc conductance.
Instead, the potential causes backscattering, and leads to lower dc
conductances.  Hence it is not unexpected that the dc conductance is
found to exhibit only gate-voltage-suppressed and not
gate-voltage-assisted feature\cite{ta6}. As the chemical potential
$\mu$ increases, the suppressed feature in the dc conductance is
characterized by dip structures at which $\mu$ is $m\hbar\omega$
above the threshold energy of a  subband.  These dip structures are
associated with the formation of QBS at a subband bottom in the
narrow channel due partly to the singular density of states (DOS).
It is interesting to see whether such QBS feature persists in
systems that have large but not singular DOS, and to explore cases
that may have gate-voltage-assisted feature.

These questions motivate us to study a saddle-point constriction
in the presence of a time-modulated gate-voltage.  A few interesting
issues are addressed.
First of all, the effective DOS in a saddle-point constriction is
not singular.  This is because the
singularity in the DOS of a narrow channel
comes from both the one-dimensionality and
the sharp threshold energy of each subband.
In a saddle-point constriction, the threshold energy of each subband
is not sharp but is smeared by tunneling processes that occur near the
threshold. The robustness of the QBS
feature against the absence of a singular DOS is explored.
Second, the gate-voltage covers regions in which the
effective width of the constriction is varying.
The gate-voltage-assisted processes become possible,
which should  be sensitive to the range of the gate-voltage.
These range-dependent characteristics are studied.
Third, the system can easily be configured into an asymmetric
situation by shifting the center of the gate-voltage away from
the symmetric center of the saddle-point constriction.
The effect of this asymmetry is studied.

In Sec. II we present our method.  In Sec. III we present some
numerical examples.  A conclusion is presented in Sec. IV.

\section{Theory}

Choosing the energy unit $E^{*}=\hbar^{2}k_{F}^{2}/2m^{*}$, the
length unit $a^{*}=1/k_{F}$, the time unit $t^{*}=\hbar/E^{*}$, and
$V_{0}$ in units of $E^{*}$, the dimensionless Schr\"{o}dinger
equation for such saddle-point constriction becomes
\begin{equation}
i\frac{\partial}{\partial t}\,\Psi(\vec{x},t) = \left[ -\nabla^{2}
 -\omega_{x}^{2}x^{2}+\omega_{y}^{2}y^{2}
 +V(x,t)\right]\,\Psi(\vec{x},t).
\end{equation}
Here $k_{F}$ is a typical Fermi wave vector of the reservoir and
$m^{*}$ is the effective mass.  The transverse energy levels
$\varepsilon_{n}=(2n+1)\omega_{y}$ are quantized, with $\phi_{n}(y)$
being the corresponding wave functions.  The time-modulated
gate-voltage $V(x,t)$ is given by
\begin{equation}
V(x,t) = V_{0} \ \theta \left({a\over 2}-\left|x-x_{c}\right|\right)\,
\cos \left( \omega t \right),
\end{equation}
where the interaction region is centered at $x_{c}$ and with a
longitudinal range $a$.  Even though the saddle-point constriction
is symmetric,  the transport characteristics could become asymmetric
if the interaction region were not centered at the symmetric center
of the constriction. On the other hand, the $V(x,t)$ we considered
is uniform in the transverse direction and it does not induce
intersubband transitions.  Thus for a $n$th subband electron, and
with energy $\mu$, incident along $\hat{x}$, the subband index $n$
remains unchanged and the scattering wave function can be written in
the form $\Psi_{n}^{+}(\vec{x},t)=\phi_{n}(y)\psi(x,t)$.

The wave function $\psi(x,t)$ can be expressed in terms of the
unperturbed wave functions $\psi(x,\mu_{n})$ which satisfy the
Schr\"{o}dinger equation
\begin{equation}\label{SE}
\left[-\frac{\partial^{2}}{\partial x^{2}}-\omega_{x}^{2}x^{2}
\right]\psi(x,\mu_{n})=\mu_{n}\psi(x,\mu_{n}),
\end{equation}
where $\mu_{n}=\mu-\varepsilon_{n}$ is the energy for the motion
along $\hat{x}$.  The solutions to \eq{SE} are doubly degenerate,
given by \cite{chu4}
\begin{mathletters}
\begin{eqnarray}
\psi_{{\rm e}}(x,\mu_{n})&=&{\rm exp}\left(\frac{-i\omega_{x}x^{2}}{2}
                                               \right)     \nonumber \\
& &\times M\left(\frac{1}{4}+i\frac{\mu_{n}}{4\omega_{x}},\frac{1}{2},
   i\omega_{x}x^{2}\right),
\end{eqnarray}
and
\begin{eqnarray}
\psi_{{\rm o}}(x,\mu_{n})&=&x\sqrt{\omega_{x}}\,{\rm exp}\left(
\frac{-i\omega_{x}x^{2}}{2}\right)  \nonumber \\
& &\times M\left(\frac{3}{4}+i\frac{\mu_{n}}{4\omega_{x}},
\frac{3}{2},i\omega_{x}x^{2}\right).
\end{eqnarray}
\end{mathletters}

Here, $\psi_{{\rm e}}\,(\psi_{{\rm o}})$ is an even (odd) function
of $x$, and $M(a,b,z)$ is the Kummer function\cite{AB}.  For our
scattering problem, it is more convenient to construct out of
$\psi_{{\rm e}}$ and $\psi_{{\rm o}}$ wave functions that have the
appropriate asymptotic behaviors.  In the asymptotic region $x\to
-\infty$, we construct wave functions $\psi_{{\rm in}}$ and
$\psi_{{\rm ref}}$ which have only positive and negative current,
respectively.  In the asymptotic region $x\to +\infty$, we construct
wave function $\psi_{{\rm tran}}$ which has only positive current.
These wave functions are given by\cite{chu4}
\begin{mathletters}
\begin{equation}
\psi_{{\rm in}}(x,\mu_{n})=\psi_{{\rm o}}(x,\mu_{n})+\alpha(\mu_{n})
\,\psi_{{\rm e}}(x,\mu_{n}),
\end{equation}
\begin{equation}
\psi_{{\rm ref}}(x,\mu_{n})=\psi_{{\rm o}}(x,\mu_{n})
+\alpha(\mu_{n})^{*}\,\psi_{{\rm e}}(x,\mu_{n}),
\end{equation}
and
\begin{equation}
\psi_{{\rm tran}}(x,\mu_{n})=\psi_{{\rm o}}(x,\mu_{n})-
\alpha(\mu_{n})^{*}\,\psi_{{\rm e}}(x,\mu_{n}),
\end{equation}
\end{mathletters}
where
\begin{eqnarray}
\alpha(\mu_{n}) &=& \frac{1}{4\pi}\left[{\rm exp}\left(
\frac{-\pi\mu_{n}}{4\omega_{x}}\right)-i\,{\rm exp}\left(
\frac{\pi\mu_{n}}{4\omega_{x}}\right)\right] \nonumber \\
& & \times\left|\,\Gamma\left(\frac{1}{4}+i\frac{\mu_{n}}{4\omega_{x}}
\right)\right|^{2},
\end{eqnarray}
and $\Gamma(z)$ is the Gamma function.

Using these wave functions, the wave function $\psi(x,t)$ that
corresponds to an electron incident from the left hand side of the
constriction can be written in the form\cite{C5,ta6}
\begin{widetext}
\begin{eqnarray}\label{wavefun}
\psi(x,t) &=&
 \psi_{{\rm in}}(x,\mu_{n})e^{-i\mu t} + {\displaystyle \sum_{m}}\,
   r_{m}^{(+)}(\mu_{n},x_{c})\,
   \psi_{{\rm ref}}(x,\mu_{n}+m\omega)\,e^{-i(\mu+m\omega)t},
   \ \ {\rm if}\ x < x_{0} \nonumber \\
 \psi(x,t) &=&{\displaystyle \int}\,d\epsilon \ \left[
   \tilde{A}(\epsilon)\psi_{{\rm ref}}(x,\epsilon-\varepsilon_{n})
   +\tilde{B}(\epsilon)\psi_{{\rm tran}}(x,\epsilon-
   \varepsilon_{n})\right]
   \,e^{-i\epsilon t} \nonumber \\
   &&\hspace{7 mm}\times\,{\displaystyle \sum_{p}}\,\left[
   {\displaystyle
   J_{p}\left({V_{0}\over \omega} \right)\,e^{-ip\omega t}
   }
   \right], \ \ {\rm if}\ x_{0} < x < x_{1} \\
 \psi(x,t) &=& {\displaystyle \sum_{m}}\,t_{m}^{(+)}(\mu_{n},x_{c})\,\psi_{{\rm tran}}
   (x, \mu_{n}+m\omega)\,
   e^{-i(\mu+m\omega)t},
   \ \ {\rm if}\ x > x_{1} \nonumber
\end{eqnarray}
\end{widetext}
where $n$ is the subband index, $m$ is the sideband index, and
$J_{p}(x)$ is the Bessel function.  The superscript $(+)$ in the
transmission and the reflection coefficients indicates that the
electron is incident from the left hand side of the constriction.
The sideband index $m$ corresponds to a net energy change of
$m\hbar\omega$ for the outgoing electrons.  The two ends of the
interaction region are at $x_{0}=x_{c}-a/2$ and $x_{1}=x_{c}+a/2$.

The transmission and the reflection coefficients can be obtained
from matching the wave functions and their derivatives at the two
ends of the time-modulated gate-voltage.  For the matching to hold
at all times, the integration variable $\epsilon$ in \eq{wavefun}
has to take on discrete values $\mu\pm m\omega$.  Hence we can write
$\tilde{A}(\epsilon)$ and $\tilde{B}(\epsilon)$ in the form
\begin{equation}
\tilde{F}(\epsilon) = \sum_{m} \, F(m)\,
                        \delta(\epsilon-\mu-m\omega),
\end{equation}
where $\tilde{F}(\epsilon)$ refers to either
$\tilde{A}(\epsilon)$ or $\tilde{B}(\epsilon)$.  After
performing the matching and eliminating the reflection
coefficients $r_{m}^{(+)}(\mu_{n},x_{c})$, we obtain the equations
relating $A(m),\, B(m)$ and the transmission coefficients
$t_{m}^{(+)}(\mu_{n},x_{c})$, given by
\begin{widetext}
\begin{eqnarray}
\psi_{{\rm tran}}(x_{1},\mu_{n}+m\omega)\,t_{m}^{(+)}(\mu_{n},x_{c})
   &=& \sum_{m'}\,\left[\,A(m')\,\psi_{{\rm ref}}(x_{1},\mu_{n}
     +m'\omega)+B(m')\,\psi_{{\rm tran}}(x_{1},\mu_{n}+m'\omega)
     \right] \nonumber \\
   & & \hspace{7mm}\times\,J_{m-m'}\left( {V_{0}\over \omega} \right),
\label{BC1}
\end{eqnarray}
\begin{eqnarray}
\psi_{{\rm tran}}'(x_{1},\mu_{n}+m\omega)\,t_{m}^{(+)}(\mu_{n},x_{c})
   &=& \sum_{m'}\,\left[\,A(m')\,\psi_{{\rm ref}}'
       (x_{1},\mu_{n}+m'\omega)+B(m')\,\psi_{{\rm tran}}'
       (x_{1},\mu_{n}+m'\omega)\right] \nonumber \\
   & & \hspace{7mm}\times\,J_{m-m'}\left( {V_{0}\over \omega} \right),
\label{BC2}
\end{eqnarray}
and
\begin{eqnarray}
& & \left[\, \psi_{{\rm in}}(x_{0},\mu_{n})\,\psi_{{\rm ref}}'
(x_{0},\mu_{n})-\psi_{{\rm in}}'(x_{0},\mu_{n})\,
\psi_{{\rm ref}}(x_{0},\mu_{n}) \right]\,\delta_{m0} \nonumber \\
& & \nonumber \\
&=& \ \ \sum_{m'}\,\left[\, \psi_{{\rm ref}}(x_{0},\mu_{n}+m'\omega)\,
     \psi_{{\rm ref}}'(x_{0},\mu_{n}+m\omega)-\psi_{{\rm ref}}'
     (x_{0},\mu_{n}+m'\omega)\,\psi_{{\rm ref}}(x_{0},\mu_{n}
     +m\omega)\, \right] \nonumber \\
& & \hspace{7mm}
     \times\,A(m')\,J_{m-m'}\left( {V_{0}\over \omega} \right)
     \nonumber \\
& &  \nonumber \\
& & +\sum_{m'}\,\left[\, \psi_{{\rm tran}}(x_{0},\mu_{n}+m'\omega)\,
     \psi_{{\rm ref}}'(x_{0},\mu_{n}+m\omega)-\psi_{{\rm tran}}'
     (x_{0},\mu_{n}+m'\omega)\,\psi_{{\rm ref}}(x_{0},\mu_{n}
     +m\omega)\, \right] \nonumber \\
& &  \hspace{7mm}
      \times\,B(m')\,J_{m-m'}\left( {V_{0}\over \omega}\, \right),
\label{BC3}
\end{eqnarray}
\end{widetext}
where $\psi ' = \partial\psi/\partial x$.

Solving Eqs. (\ref{BC1})-(\ref{BC3}), we obtain
$t_{m}^{(+)}(\mu_{n},x_{c}), A(m),$ and $B(m)$, from which the
reflection coefficients $r_{m}^{(+)}(\mu_{n},x_{c})$ can be
calculated.  The corresponding coefficients for electrons incident
from the right hand side of the constriction can be found following
similar procedure. The correctness of the transmission and the
reflection coefficients can be checked by a conservation of current
condition, given by
\begin{eqnarray}
\displaystyle&&{\sum_{m}}
\left|\frac{\Gamma\left(\displaystyle{\frac{1}{4}+i\,
\frac{\mu_{n}+m\omega} {4\omega_{x}}} \right)}
{\Gamma\left(\displaystyle{\frac{1}{4}+i\, \frac{\mu_{n}}
{4\omega_{x}}} \right)}\right|^{2}\, {\rm exp}\left(\frac{\pi
m\omega}{4\omega_{x}}\right)\,
 \nonumber \\
&& \times \left[\,|t_{m}^{(\sigma)}(\mu_{n},x_{c})|^{2}+
|r_{m}^{(\sigma)}(\mu_{n},x_{c})|^{2}\right] = 1,
\end{eqnarray}
where the superscript $\sigma=\pm 1$ indicates the direction
of the incident particle.  In our calculation, a large enough
cutoff to the sideband index is imposed.  The
$r_{m}^{(\sigma)}(\mu_{n},x_{c}), t_{m}^{(\sigma)}(\mu_{n},x_{c})$
coefficients that we obtain are exact in the numerical
sense.

The current transmission coefficient $T_{nm}^{\sigma}(E,x_{c})$
is the ratio between the transmitting current in the $mth$
sideband and the corresponding incident current due to a
$n$th subband electron, with incident energy $E$, and
incident direction $\sigma$.  This current
transmission coefficient is related to the transmission
coefficient, given by
\begin{eqnarray}
T_{nm}^{\sigma}(E,x_{c}) &=& \,|t_{m}^{(\sigma)}(E_{n},x_{c})|^{2} \
  {\rm exp}\left(\displaystyle{\frac{\pi m\omega}
                                    {4\omega_{x}}}\right)\,
  \nonumber \\
  & &\hspace{4mm}\times\,\left|\,\frac{\Gamma\left(
   \displaystyle{\frac{1}{4}+i\,\frac{E_{n}+m\omega}{4\omega_{x}}}
   \right)}{\Gamma\left(
   \displaystyle{\frac{1}{4}+i\,\frac{E_{n}}{4\omega_{x}}}
   \right)}\right|^{2},
\end{eqnarray}
where $E_{n}=E-\varepsilon_{n}$.  The total current
transmission coefficient $T^{\sigma}(E,x_{c})$ is defined as
\begin{equation}
T^{\sigma}(E,x_{c}) = \sum_{n}\,T_{n}^{\sigma}(E,x_{c})
 = \sum_{n}\sum_{m}\,T_{nm}^{\sigma}(E,x_{c}).
\end{equation}
Furthermore, in the case when the saddle-point potential is
shifted by $\Delta U$, the total current transmission coefficient
becomes $T^{\sigma}(E-\Delta U,x_{c})$.

We find that the total current transmission coefficients
$T^{+}(E,x_{c})$ and $T^{-}(E,x_{c})$ are different for $x_{c}\ne 0$,
when the interaction region is not centered at the symmetric center of
the constriction. This can be understood from
the following example, when the entire interaction
region is, say, on the right hand side of the constriction and the
incident energy $E$
is chosen such that the electrons have to tunnel through the
constriction.
In this example, an electron incident from the left hand side receives
no
assistance from the time-modulated gate-voltage when tunneling through
the constriction.   Rather, the electron suffers additional
reflection from the gate-voltage after tunneling through the
constriction.
However, for an electron incident from the right hand side, it can
receive assistance from the gate-voltage when passing through the
constriction.  Of course, the electron might be reflected by this
gate-voltage as well.  But in the opening up of a new
gate-voltage-assisted
transmission channel, when the electron, after absorbing $m\hbar\omega$,
can propagate, rather than tunneling, through the constriction, the
assisted
feature dominates.  This example, though not a generic one,
illustrates that the difference between the current transmission
coefficients originates from the different extent the time-modulated
gate-voltage involves in assisting the transmitting electrons.

The fact that $T^{+}(E,x_{c})$ can be different from $T^{-}(E,x_{c})$,
when the QPC is acted upon by a time-modulated potential, leads to a
nonzero current in an unbiased QPC.  The current is the photocurrent
$I_{\rm ph}$ (see below).  Therefore, the transport in the QPC
is better represented by the differential conductance $G$, rather than
the conductance, or the total current transmission coefficients $T$.

To find the differential conductance in the low-bias regime, we
choose the left reservoir to be the source electrode  such that the
left reseroir has a chemical potential shift of
$(1-\beta)\Delta\mu$,  and the right reservoir has a chemical
potential shift of $-\beta\Delta\mu$. In the low-bias regime, we
have $\Delta\mu\ll \mu$. The parameter $\beta$ had been adopted by
Martin-Moreno {\it et al.\/}\cite{m92} and Ouchterlony {\it et
al.\/}\cite{o95} in their work on the nonlinear dc transport through
a saddle-point constriction. The current $I$ in the constriction is
then given by\cite{fnte1}
\begin{eqnarray}\label{I(T)}
I &=& -\frac{2e}{h}\,\int_{-\infty}^{\infty}\,dE\,\left[ \,
      f\left( E-\mu-(1-\beta)\Delta\mu \right)\,
      T^{+}\left( E,x_{c}\right)
      \right.
      \nonumber \\
      &&\left. \mbox{\hspace {15 mm}} - \,f\left(E-\mu+\beta\Delta\mu
      \right)\, T^{-}\left( E,x_{c} \right)
      \right],
\end{eqnarray}
\noindent where $f(E)=[1+{\rm exp}(E/k_{B}T)]^{-1}$ is the Fermi
function. Here $-e$ is the charge of an electron. Assuming that the
lowest energy electrons from the reservoirs contribute negligibly to
$I$, we can extend the lower energy limit of the above integral to
$-\infty$.  The zero temperature limit of \eq{I(T)} is given by
\begin{eqnarray}\label{I(0)}
I &=& -\frac{2e}{h}\,\left[
     \int_{-\infty}^{\mu+(1-\beta)\Delta\mu}\,dE \
      T^{+}\left( E,x_{c} \right)\right.\nonumber \\
     &&- \left.\int_{-\infty}^{\mu-\beta\Delta\mu}\,dE \
      T^{-} \left(E,x_{c} \right)
      \right].
\end{eqnarray}
The differential conductance in the low bias regime, defined by
\begin{equation}
G_{0} = \left.\frac{\partial I}{\partial V_{\rm sd}}\,\right|_{V_{\rm
sd}=0}\, ,
\end{equation}
can be calculated from differentiating \eq{I(0)}, and is given by
\begin{equation}
G_{0} = \frac{2e^2}{h}\,
        \left[\,T^{+}(\mu,x_{c})(1-\beta) + T^{-}(\mu,x_{c}) \beta
        \,\right] \, .
\end{equation}
It is interesting to note that $G_{0}$ depends on $\beta$
whenever $T^{+}(\mu,x_{c})\ne T^{-}(\mu,x_{c})$,
and that this $\beta$-dependence in
$G_{0}$ does not occur for the cases of purely elastic scatterings,
such as impurity scatterings.
We  expect $G_{0}$ to be the major
contribution to the differential conductance $G$.  The other
contribution to $G$ is from the change in the
photocurrent $I_{\rm ph}$ when the QPC is subjected to the low-biased
transport field.  This term is much smaller than $G_{0}$ and is
qualitatively given by
\begin{equation}
G_{\rm ph} = { - e \over 2\omega_{x}^{2}L }\,{\partial \over \partial
x_{c}}
             I_{\rm ph}(\mu,x_{c}) \, ,
\end{equation}
where
\begin{equation}
 I_{\rm ph}(\mu,x_{c}) = -{2e \over h}\int_{-\infty}^{\mu}\, dE \left[
              T^{+}(E,x_{c}) -  T^{-}(E,x_{c}) \right]
\end{equation}
is the photocurrent.
Here $L$ is the effective length of the potential drop across the QPC,
and $\Delta\mu/(2\omega_{x}^{2}L)$ is the effective shift of the QPC
position caused by the small bias potential.
The differential conductance $G = G_{0} + G_{\rm ph}$.

\section{Numerical examples}

In our numerical examples, the physical parameters are taken to be
that in a high-mobility ${\rm GaAs-Al_{x}Ga_{1-x}As}$
heterostructure, with a typical electron density
$n\sim 2.5\times 10^{11}$ cm$^{-2}$
and $m^{*}=0.067m_{e}$.  Correspondingly, we choose an energy unit
$E^{*}=\hbar^{2}k_{F}^{2}/(2m^{*})=9$ meV, a length unit
$a^{*}=1/k_{F}=79.6$ \AA, and a frequency unit
$\omega^{*}=E^{*}/\hbar=13.6$ THz.  For the saddle-point
constriction, we have chosen $\omega_{x}=0.0125$, and
$\omega_{y}=0.05$ such that the effective length to width ratio of
the constriction is $L_{c}/W_{c}=\omega_{y}/\omega_{x}=4$.
 In presenting the dependence of $G$ on $\mu$, it is
more convenient to plot $G$ as a function of $X$, where
\begin{equation}
 X={1\over 2} \left[ {\mu \over \omega_{y}} + 1 \right].
\end{equation}
The integral value of $X$ is the number of propagating channels
through the constriction.

To evaluate the term $G_{\rm ph}$ of the differential conductance
$G$, the length $L$ of the potential drop  is taken to be of the
same order as $L_{c}$\cite{lev89},  where the length of our
constriction  $L_{c}=4W_{c}=8\sqrt{3/\omega_{y}}\simeq 62$ for a
typical $n=1$ subband.  Following Ouchterlony {\it et al.\/}, who
have chosen $L_{c}/L\approx 1.5$\cite{fnte2}, we choose $L$ to have
a value $L = L_{c}/1.5\simeq 41$.

In Figs.\ 2, 3, and 4, we present the changes in the $G$
characteristics when the range of the time-modulated potential is
increased, from
$a = 16,\ 32, {\rm to}\ 50$, respectively.  All these time-modulated
potentials are
centered, with $x_{c}=0$, and have the same frequency ($\omega=0.04$),
and
the same amplitude ($V_{0}=0.06$).  The bias parameter $\beta=0.5$ in
these
figures.  The $G$ characteristics are represented by the dependence of
$G$ on $X$, the suitably rescaled chemical potential $\mu$.  According
to this scale, when $\mu$ is changed by a subband energy spacing,
it corresponds to $\Delta X=1$, and when $\mu$ is changed by
$\hbar\omega$, it corresponds to $\Delta X=\omega/(2\omega_{y})=0.4$.
In addition, when $X=N$,
$\mu$ is at the threshold of the $N$th subband.

\begin{figure}[btp]
      \includegraphics[width=0.40\textwidth,angle=0]{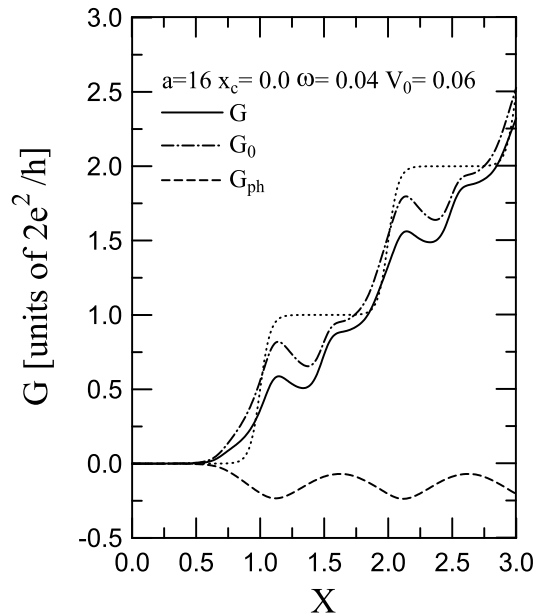}
\caption{Differential conductance $G$ as a function of $X$
           for a centered time-modulated potential ($x_{c}=0$), with
           oscillating amplitude $V_{0}=0.06$, frequency $\omega=0.04$,
           and $\beta=0.5$.  The range of the  potential $a=16$
           covers a
           distance up to $d = 8$ from the constriction center.
           The solid curve is the total differential conductance $G$,
           the
           dashed-dotted curve is $G_{0}$, and the dashed curve is
           $G_{\rm ph}$.  In the assisted regime, $G_{\rm ph}$
           suppresses
           the differential conductance, but the general
           tunneling-like
           feature remains unchanged.  In the suppressed regime, there
           are dip structures at $X=1.4, \,{\rm and} \,2.4$.}
\end{figure}
In Fig.\ 2, we find both gate-voltage-assisted and
gate-voltage-suppressed features in $G$.  These two features occur
in well separated regions of $X$. The gate-voltage-assisted regions
occur when $\mu$ is just beneath a subband threshold, and is most
evident in the pinch-off ($X<1$) region, while the
gate-voltage-suppressed regions occur when $\mu$ is above but close
to a subband threshold.  Dip structures are found in the suppressed
region, at around $X=1.4,\, {\rm and} \ 2.4$, that is, at $\Delta
X=0.4$ above a threshold.  These dip structures are due to the
processes that an electron in the $N$th subband, and at energy
$N+\Delta X$, can give away an energy $\hbar\omega$ and become
trapped in the quasi-bound-state (QBS) just beneath the
threshold\cite{b2,ta6}.  In contrast with the QBS features in narrow
channels\cite{ta6}, the QBS structures in a saddle-point contriction
is much broader, indicating that the QBS life-time is much shorter
due to the added possibility of escape via tunneling. In the
gate-voltage-assisted region, $G$ increases gradually, rather than
abruptly, when a channel, after picking up an energy $\hbar\omega$,
becomes propagating.  This is because the range of the interacting
region does not cover far enough so that the electron, though having
the right energy, has to tunnel to the interacting region first
before being assisted.  This also explains why there are no
structures at $N\pm 2\Delta X$, which corresponds to the
$2\hbar\omega$ processes.  Consequently, the dip structures in the
suppressed regions are not affected by the gate-voltage-assisted
features.

It is interesting to note that there is no harmonic feature that could
have been caused by the abrupt-profile of the the gate-voltage.  This
is because the effective wavelength of a particle decreases as it
emanates from the constriction so that multiple
 scattering between the two abrupt edges of the
potential is subjected
 to rapid phase fluctuations, suppressing any possible
harmonic resonances.  Thus our results should represent also
the cases of smooth-profile gate-voltages.

\begin{figure}[btp]
      \includegraphics[width=0.40\textwidth,angle=0]{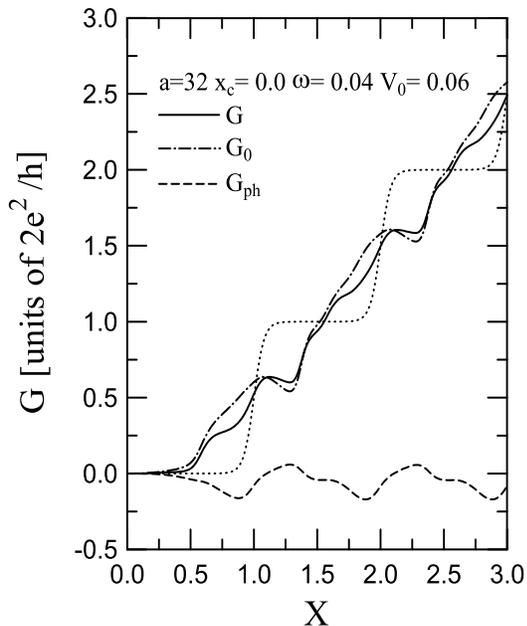}
      \caption{Differential conductance $G$ as a function of $X$
           for a centered time-modulated potential.  The physical
           parameters are the same as in Fig.\ 2 except that the
           range of
           the potential is $a=32$.  The potential covers a distance up
           to $d = 16$ from the constriction center.
           In the assisted regime, the $G_{\rm ph}$ modifies the
           shoulder-like feature in $G_{0}$ and leads to the
           quasi-mini-step-like feature in $G$.  In the suppressed
           regime, the dip structures at $X=1.4,\,{\rm and}\, 2.4$
           are slightly modified by $G_{\rm ph}$ }
\end{figure}
In Fig.\ 3, the QBS structures at $X=1.4,$ and $\,2.4$ are still
evident.  The assisted features are enhanced.  In particular, in
the pinch-off region, $G$ increases much faster, showing a
mini-step-like structure.  The gate-voltage-assisted and the
gate-voltage-suppressed features are well separated because no
$2\hbar\omega$ features are found.

\begin{figure}[btp]
      \includegraphics[width=0.40\textwidth,angle=0]{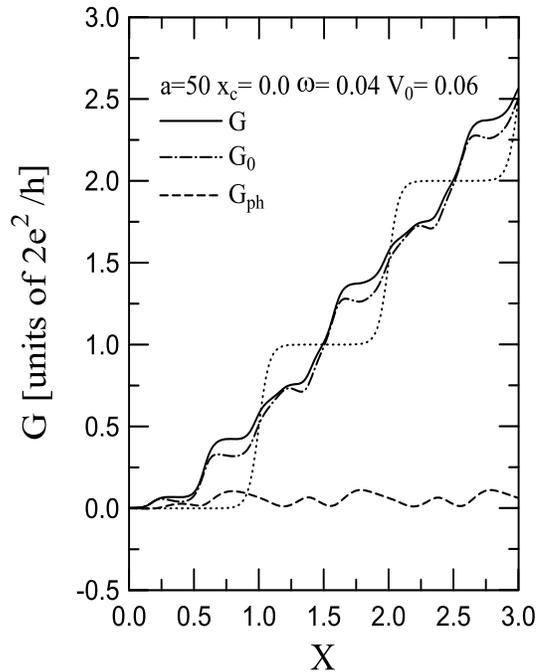}
      \caption{Differential conductance $G$ as a function of $X$
           for a centered time-modulated potential.  The physical
           parameters are the same as in Fig.\ 2 except that the range
           of the potential is $a=50$, which covers a distance up to
           $d = 25$ from the constriction center.  In the assisted
           regime,  $G_{\rm ph}$ enhances $G$.  There are two
           mini-step-like structures in the pinch-off region.}
\end{figure}

In Fig.\ 4, there are additional dip structures in $G_{0}$ at
$X=1.8, \,$ and $\,2.8$, which
 indicates that $2\hbar\omega$ processes become
significant.  There are, of course, assisted features that involve one
$\hbar\omega$ processes, and they are the abrupt rises in $G$ at
$X=0.6, 1.6,\, {\rm and}\, 2.6$.  The assisted feature that involves
$2\hbar\omega$ is most clearly demonstrated in the pinch-off region,
around $X=0.2$, where $G$ exhibits another ministep.  Other
$2\omega$ assisted
processes are at $X=1.2,\, {\rm and}\, 2.2$,
which, unfortunately, are in the
vicinity of the dip structures at $X=1.4,\, {\rm and}\, 2.4$.
Hence the dip structures become less dip-like
but have turned into a sharp uplift in $G$,
 because they are affected by the assisted features.

The assisted features in the above three figures are different, and the
difference is associated with how the electrons enter the
interaction region.  For the $N$th subband electrons with incident
energies that fall within the $m\hbar\omega$ interval below the
threshold of the same subband, they are nonpropagating.  They can become
propagating, and traverse through the constriction, by absorbing
$m\hbar\omega$ from the time-modulated potential.  But the electrons
have to
be in the interaction
region to absorb the needed energy. If the gate-voltage
covers a region over
a distance $d>d_{m}=\sqrt{m\omega}/\omega_{x}$ from the
center, and on the
incident side, of the constriction, the incident electrons
can propagate into the interaction region.  However, if the gate-voltage
only covers regions over a shorter distance ($d<d_{m}$) from the
constriction center, the electrons have to tunnel into the interaction
region.  For $\omega_{x}=0.0125$, we have $d_{1}=16$, and
$d_{2}\approx 23$.  We recall that the distances $d$ cover by the
gate-voltage
are $d=8,\ 16, {\rm and}\ 25$, respectively, in Figs.\ 3, 4, and
5.  Hence in Fig.\ 2, $d < d_{1}$, and the electrons have to
tunnel into the interaction regime so that the assisted feature, such
as in the $X\leq 1$ region,
exhibits a tunneling-like structure.  In Fig.\ 3, when $d=d_{1}$, the
electrons can barely
avoid entering the interaction region via tunneling, the
assisted feature exhibits a quasi-mini-step-like structure.  In Fig.\ 4,
when $d>d_{2}$, the electrons involving in the $2\hbar\omega$
processes can also propagate into the interaction region, and the
assisted feature exhibits additional mini-step structures.

\begin{figure}[btp]
      \includegraphics[width=0.40\textwidth,angle=0]{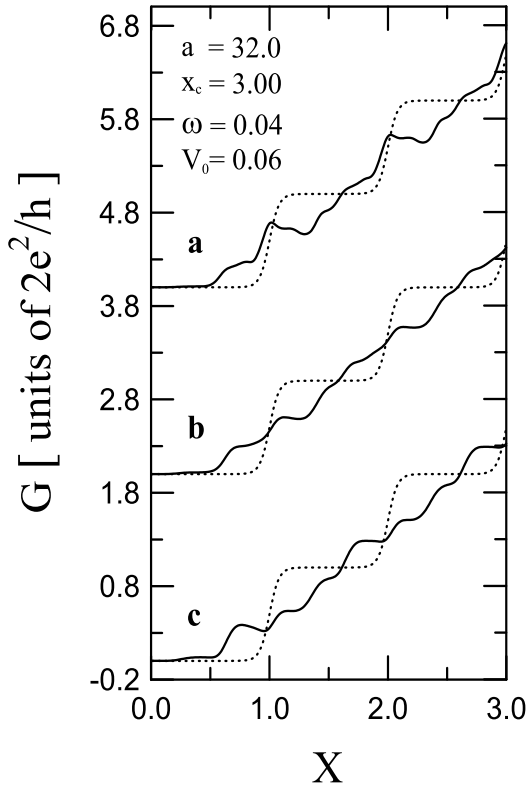}
      \caption{Differential conductance $G$ as a function of $X$ for an
           off-centered time-modulated potential ($x_{c}=3$), with range
           $a=32$, frequency $\omega= 0.04$, and oscillating amplitude
           $V_{0}=0.06$.  The parameter $\beta=0.2
           (a),\ 0.5 (b),\ {\rm and}\ 0.8 \ (c)$.  The curves are
           vertically offset for clarity.}
\end{figure}

In Fig.\ 5, we present the dependence of the $G$ characteristics on
the parameter $\beta$.  The time-modulated gate-voltage is
off-centered, with $x_{c}=3.0$, range $a=32$, $V_{0}=0.06$, and
frequency $\omega=0.04$.  The parameter $\beta=0.2,\ 0.5,\ {\rm
and}\ 0.8$ in Figs.\ 5(a), (b), and (c), respectively. The curves in
Fig.\ 5 show that $G$ is quite sensitive to $\beta$.  But from
further analysis, we find that it is $G_{0}$, rather than $G_{\rm
ph}$, that gives rise to the $\beta$-sensitivity in $G$. Hence,
according to \eq{I(T)}, as $\beta$ increases, the contribution to
$G_{0}$ from $T^{+}$ decreases while that from $T^{-}$ increases.
Since the assisted feature of $T^{-}$ is prominent than that of
$T^{+}$ because $x_{c}=3.0$, therefore the assisted features in the
pinch-off region  are enhanced progressively in Figs.\ 5(a), (b),
and (c). This particular $\beta$-dependence simply reflects the
asymmetry in $T^{+}(\mu,x_{c})$ and $T^{-}(\mu,x_{c})$, which occurs
for coherent inelastic scatterings and not for elastic scatterings.

\begin{figure}[btp]
      \includegraphics[width=0.40\textwidth,angle=0]{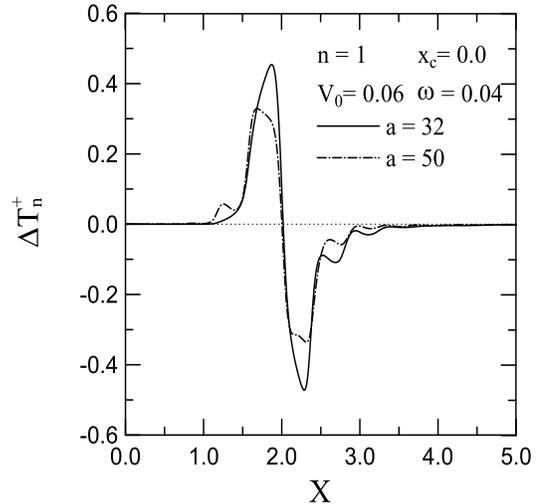}
      \caption{$\Delta T_{n}^{+}$ as a function of $X$
           for a centered time-modulated potential and subband $n=1$,
           with frequency $\omega=0.04$, $\beta=0.5$, and oscillating
           amplitude $V_{0}=0.06$.  The potential range $a = 32 \,
           ({\rm solid}), \, {\rm and} \, 50 \,
           ({\rm dashed-dotted})$.  The assisted feature is found
           below $X=2$ (the n=1 subband edge), and the suppressed
           feature is found above $X=2$.  The $2\hbar\omega$
           structures appear in the longer potential range ($a=50$)
           case.  The QBS features are at around
           $X=2.4,\ 2.8,\ {\rm and}\ 3.2$.}
\end{figure}

\section{Conclusion}

We have shown that the differential conductance $G$,
rather than the conductance
or the current transmission coefficient, is the relevant physical
quantity for the characterization
of the low-bias transport, when the QPC is acted upon
by a time-modulated field.
This has not been recognised previously.  Thus
to compare with the results
of previous studies, we can only turn to the
current transmission coefficient.  The deviation of the current
transmission coefficient from its unperturbed value
\begin{displaymath}
 \Delta T^{+}_{n}(E,x_{c})=T^{+}_{n}(E,x_{c})-{\cal T}^{0}_{n}(E),
\end{displaymath}
was the photoconductance calculated by Grincwajg {\it et
al.\/}\cite{g5}, and Maa\o\ {\it et al.\/}\cite{m6}, when they
considered a transverse electric field acting on a QPC with varying
width.  Here ${\cal T}^{0}_{n}(E)=1/[1+{\rm
exp}(-\pi\varepsilon_{n})]$, and $n$ is the subband index of the
incident electron.  In Fig.\ 6, we plot our $\Delta
T^{+}_{n}(E,x_{c})$ results against $X$.  The time-modulated
potential is centered $(x_{c}=0)$, the frequency $\omega=0.04$, the
amplitude $V_{0}=0.06$, and the incident subband index $n=1$.  The
threshold for the subband is $X=2$.  The assisted and the suppressed
features are clearly shown below and above the threshold,
respectively.  This trend is the same as the results of Refs. 8, and
10, even though the inelastic processes induced by a transverse
field is different from that by a potential. On the other hand, our
results have the added QBS features in the suppressed region, and
the added $2\omega$ feature in the assisted region for a longer
potential range $a$. Both features are the main results of this
paper.

Finally, we have demonstrated the robustness of the QBS
features.  Should these QBS features exist and remain robust in the
cases of time-modulated electric field?  We are currently
investigating this possibility.

\begin{acknowledgments}
This work was supported in part by the National Science Council of
the Republic of China through Contract No. NSC86-2112-M-009-004.
\end{acknowledgments}

\bibliographystyle{apsrev}

\begin{thebibliography}{99}
\bibitem{fnte1} One may want to use another current expression,
                which, in units of $-2e/h$, is given by
        \begin{eqnarray*}
        I &=& \int_{-\infty}^{\infty}dE\, \sum_{nm}\,\left\{
              T^{+}_{nm}\left(E,x_{c}\right)\,
              f\left[E-\mu-(1-\beta)\Delta\mu\right] \right.
              \nonumber \\
             &&\hspace{20mm} \times \left[1-f(E-\mu-m\,\omega+\beta\Delta\mu)\right] \nonumber \\
          && \hspace{15mm}-\, T^{-}_{nm}(E,x_{c})\, f(E-\mu+\beta\Delta\mu)
              \nonumber \\
          && \hspace{20mm} \times \, \left.
          \left[\,1-f(E-\mu-m\omega-(1-\beta)\Delta\mu)\right]
          \right\} .
        \end{eqnarray*}
              However, it can be shown that the two
              expressions give the same result.
\end{thebibliography}

\end{document}